\documentclass[parskip=half, bibliography=totoc, captions=tableheading]{scrartcl}
\usepackage[english]{babel}

\usepackage{amsmath}
\usepackage{amssymb}
\usepackage{mathtools}

\usepackage[
  locale=DE,
  separate-uncertainty=true,
  per-mode=symbol-or-fraction,
]{siunitx}

\usepackage[section, below]{placeins}
\usepackage[
  labelfont=bf,
  font=small,
  width=0.9\textwidth,
]{caption}
\usepackage{subcaption}
\usepackage{graphicx}

\usepackage{float}
\usepackage[
  backend=biber,
  style=ieee,
  maxnames=4
]{biblatex}
\title{Machine learning based luminance analysis of a µLED array}

\author{Steven Becker | steven.becker@tu-dortmund.de \\ \\ Experimentelle Physik 2, TU Dortmund University \\ Otto-Hahn-Straße 4a, 44227 Dortmund, Germany}

\begin{document}
\maketitle

\begin{abstract}
In the past years, the development of µLED arrays gained momentum since they combine the advantages of LEDs, such as high brightness and longevity, with the high resolution of a micro-scaled structure. For their development, spatially resolved measurement of luminance and color of single µLEDs and the entire light-emitting surface are usually analyzed as they quantify the visual perception. However, studying individual µLEDs is time-consuming to measure and evaluate, while examining the entire light-emitting area suffers from interference from non-functioning µLEDs. This paper presents a method to perform both analyzes with a single measurement employing unsupervised machine learning. The results suggest that a precise reconstruction of the µLEDs and a more accurate characterization µLED arrays is achieved.
\end{abstract}

\thispagestyle{empty}

\section*{Introduction}
Over the last decades, light-emitting diodes (LEDs) were established as a key
light source in the consumer market, such as for general lighting or automotive.
The accelerators for the far-reaching influence of LEDs are their durability, even
under harsh conditions such as extreme temperatures, and their high-efficiency and
luminance. However, for applications requiring high pixelation like displays, the
high-performance LEDs are mostly too big because research has focused on the
optimization of $1 \, \mathrm{mm}^2$ large high-performance LEDs~\cite{doi:10.1063/1.5096322,day_full-scale_2012}.

Shrinking those LEDs into the micro-scale - edge length below $100\,$ µ$\mathrm{m}$- allows to overcome this issue and should theoretically increase their efficiency. Already in 2000, Jin et al. published
the first realizations of micro light-emitting diodes µLED~\cite{jin_gan_2000}, and since then, numerous publications regarding
this topic were submitted.

One advantage of µLEDs is the possibility of combining
them to an array-structure, which can then be linked with an underlying electrical
control unit. This two-dimensional stringing-together of µLEDs results
in a µLED array. Such structures should have a higher illuminance and homogeneity than a single µLED, as well as
provide a high brightness, contrast, resolution, and durability~\cite{day_iii-nitride_2011, tian_effects_2017}.
In combination, they may become superior against already established pixelated light sources such as organic LEDs (OLEDs) or liquid crystal (LC) based ones~\cite{8811753, day_full-scale_2012}. Furthermore, LED arrays gained much momentum in the industry as well~\cite{8811753}.
However, there is no published data that the established manufacturing techniques, mass transfer-based or the monolithic-based fabrication, achieved a yield of close to $100\%$~\cite{review_progress}. Consequently, most of the produced µLED arrays will contain non-emitting and therefore defect µLEDs.

Classical approaches to characterize the luminance of the light-emitting surface (LES), such as averaging over the entire area,
do not distinguish between functional and defect, resulting in an underestimation of the actual µLED array behavior.
If a significant fraction of nonfunctional µLEDs occurs, this becomes a problem in the development process.
For instance, the "noise" created by the defect µLEDs could prevent an evaluation of a design change.
Each µLED (pixel) should be classified, and defect ones may not be considered for the final analysis to overcome this blurring effect.

In this paper, a machine learning-based algorithm is applied to overcome this issue. Therefore, individual µLEDs are located in a luminance image of a µLED array and
then classified as functional or defect with an unsupervised learner (KMeans). As a result, the behavior of the µLED array is studied without the blurring effect of the
non-emitting µLEDs. The implemented analysis is evaluated by three performance measures including the reconstructed µLED size, the confusion matrix of the underlying classifier, and the noise
reduction performance. In order to achieve this, a luminance camera with a resolution of $2448\times 2050$ pixels and a white light-emitting µLED array with a total of $60\times 60$ µLEDs are used. Each µLED of the light source has a size of $40 \,$ µ$\mathrm{m} \times 40\,$ µ$\mathrm{m}$ and can be addressed separately.
A lens with a magnification factor of two is mounted on the camera.
Due to the luminance camera pixel's size of $3.45\,$ µ$\mathrm{m}$, a single µLED is represented by approximately $23$ camera pixels.

\section*{Results}
\begin{figure}[h!]
  \centering
  \includegraphics[width=\textwidth]{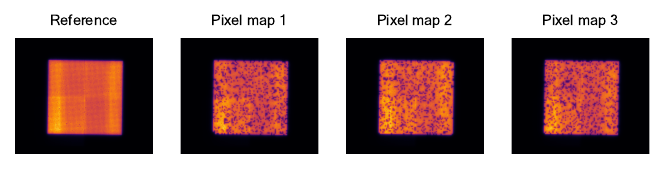}
  \caption{Luminance images of the used \emph{Pixel maps} to benchmark the analysis pipeline.}
  \label{fig: reference_images_array_level_benchmark}
\end{figure}

The developed analysis is tested with four luminance images, which are shown in Figure \ref{fig: reference_images_array_level_benchmark}.
The first image represents the \emph{Reference state}, where all µLEDs of the array are turned on. The non-uniform luminance distribution originates from the underlying electronics
and will not be discussed further. On the other images (\emph{Pixel map} one to three), some µLEDs are turned-off on purpose to simulate a nonfunctional state. These pseudo-defect
µLEDs were selected for each \emph{Pixel map} randomly and represent a yield of $80\%$, hence $20\%$ of the µLEDs are turned-off. Furthermore, the four luminance images are
the direct outputs of the luminance camera, which explains the black bars framing the array structure. In addition, the sample was rotated to increase the difficulty and to stress the robustness of the method.

\begin{figure}[h!]
  \centering
  \includegraphics[width=\textwidth]{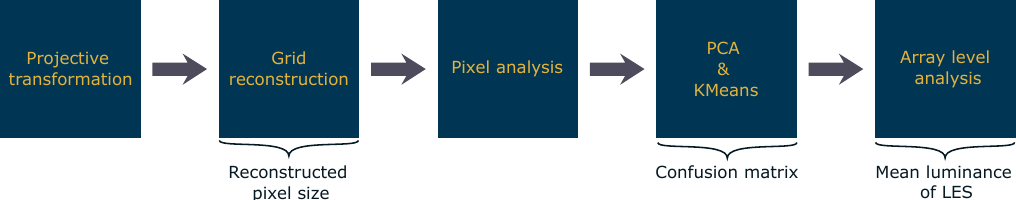}
  \caption{Schematic illustration of the analysis pipeline.}
  \label{fig: array_level_analyis_pipeline_schematical}
\end{figure}

The \emph{Pixel maps} shown in Figure \ref{fig: reference_images_array_level_benchmark} are the inputs of the underlying analysis pipeline, which is illustrated in
Figure \ref{fig: array_level_analyis_pipeline_schematical}.
The first step of the pipeline performs a \emph{projective transformation} to compensate for a tilted and rotated sample.
Then, the \emph{grid reconstruction} scopes to reassemble the array structure, so that individual µLEDs can be localized. This result is exploited by the \emph{Pixel analysis} that
studies the luminance and CIE color coordinates of each reconstructed pixel. The acquired data set is then guided into the machine learning part of the pipeline, which utilizes a \emph{principal component analysis} (PCA) and an unsupervised learner (KMeans) to classify functional and nonfunctional µLEDs. Finally, the \emph{Array level analysis} evaluates the entire light-emitting surface (LES), while considering the information about the functionality of the individual µLEDs. Note, each step of the pipeline will be discussed in more detail in the subsequent section.
In total, three performance measures can be extracted from the analysis pipeline (see Fig. \ref{fig: array_level_analyis_pipeline_schematical}):
Firstly, the reconstructed pixel size, secondly the confusion matrix of the pixel classification, and finally the mean luminance of the LES.

\begin{figure}[h!]
  \centering
  \includegraphics[width=0.6\textwidth]{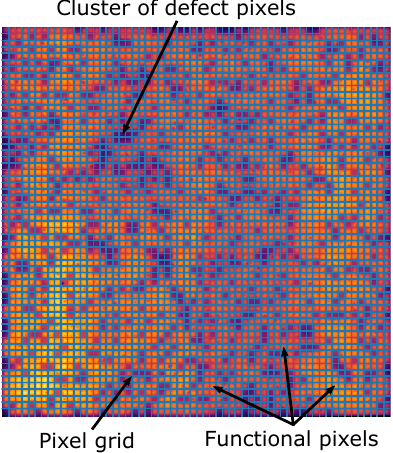}
  \caption{Reconstructed pixel grid (blue lines) of the \emph{Pixel map 1} (see Figure \ref{fig: reference_images_array_level_benchmark}).}
  \label{fig: reconstructed_pixel_grid_for_pixel_mathrm_1}
\end{figure}

Since the luminance image provides spatial information of the µLED array, the pixel grid framing the individual µLEDs can be extracted. Figure \ref{fig: reconstructed_pixel_grid_for_pixel_mathrm_1} shows the result of the underlying grid reconstruction algorithm for \emph{Pixel map 1}. In addition, Figure \ref{fig: reconstructed_pixel_grid_for_pixel_mathrm_1} confirms that the applied projective transformation can correct the initial rotation of the sample.
The method is even capable of reconstructing the grid at locations where several pixels are "defect". The quality of the extracted grid can be measured by its pixel size since the pixel size of the grid ideally corresponds to the number of camera pixels representing a µLED.
Equation \eqref{eq: reconstructed_pixel_size} shows the mean grid's pixel size $\overline{d}$ for all three \emph{Pixel maps} (compare Figure \ref{fig: reference_images_array_level_benchmark}):
\begin{equation}
  \label{eq: reconstructed_pixel_size}
  \overline{d}_{\mathrm {pixel, 1}} = \overline{d}_{\mathrm {pixel, 2}} = \overline{d}_{\mathrm {pixel, 3}} = \left(23\times23\right)\,\mathrm{px}^2.
\end{equation}
The standard deviation of the mean value is intentionally neglected
because the uncertainty is less than $0.5\,\mathrm{px}$. As a result, it is concluded that the pixel grid reconstruction is capable of consistently reconstructing
the proposed pixel size. Note that pixels at the edges of the LES are currently not considered to diminish the influence of boundary effects.

\begin{figure}[!h]
  \centering
  \includegraphics[width=\textwidth]{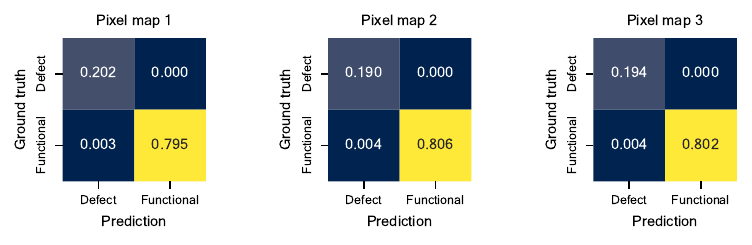}
  \caption{Confusion matrices for the pixel classification of the three \emph{Pixel maps}. The pipeline achieves an accuracy of $\approx 99.5\%$ for each \emph{Pixel map},
  as indicated through the diagonal elements. Further, the lower left and upper right matrix elements reveal a false negative rate of under $0.5\%$ and a false positive rate of $0\%$, respectively.}
  \label{fig: confusion_matrix_use_case}
\end{figure}

Next, the reconstructed grid organizes the input luminance image (compare Figure \ref{fig: reference_images_array_level_benchmark}) into
individual pixel areas containing information about a certain µLED. Hence, from a single luminance image, information about hundreds of
µLEDs can be extracted and used for statistical analysis. The maximum, mean, minimum, and standard deviation of the
luminance as well as the mean $\mathrm{CIE} \, x$  and $\mathrm{CIE} \, y$ color coordinates are calculated for each pixel including the
"defect" ones. Consequently, this procedure leads to a statistical data set that not only represents a base for a profound analysis of the
µLEDs behavior but also serves here as training data for the unsupervised learner.
In particular, the classification of defect pixels is of interest as elaborated in the introduction. The achieved performance of the used \emph{k-Means-algorithm} is illustrated in the confusion matrices shown in Figure \ref{fig: confusion_matrix_use_case}. The confusion matrices reveal for each \emph{Pixel map} (compare Figure \ref{fig: reference_images_array_level_benchmark}) the corresponding prediction accuracy for the classification \emph{functional}\, /\, \emph{defect}. Note that the general procedure of the classification step is explained in detail in the subsequent section.
According to the matrices, the classifier predicts the pixel's status with a
percentage of $\approx 99.5\%$ correct. In the other case, the lower left matrix elements indicate that functional pixels
are falsely classified as defective in less than $0.5\%$. Moreover, the obtained true positive and negative rates of about $80\%$ and $20\%$,
respectively, are consistent with the originally selected yield of $80\%$. In general, from the confusion matrix
further evaluation quantities can be extracted including, but not limited to, the
precision score $P$, recall score $R$, and the $F1$-score. The precision score measures
the ability of a classifier to prevent the positive labeling of something that should be negative. In the case of this classifier, it quantifies the risk of
predicting a defect pixel as functional. The recall score represents the capability of a classifier to identify all positive elements, here the ability to notice functional pixels. The $F1$-score considers both precision and recall score to give a balanced quantity~\cite{tharwat_classification_2020}. Note, for all scores $0$ is the worst and $1$ is the best outcome. The scores can be calculated as follows:
\begin{align}
 \begin{aligned}
    TP = &\mathrm{True \,positiv}, \qquad FP = \mathrm{False \, positiv}, \qquad  FN = \mathrm{False \, negative} \\
    P &= \frac{TP}{TP+FP}, \qquad R = \frac{TP}{TP+FN}, \qquad F1 = 2\frac{P\cdot R}{P + R}.
    \end{aligned}
\end{align}
The calculation of those measures confirms the performance of the \emph{k-Means}-based classifer:
\begin{align}
    P_1 &= P_2 = P_3 \approx 0.995 \approx 1 \\
    R_1 &= R_2 = R_3 = 1 \\
    F1_1 &= F1_2 = F1_3 = 1.
\end{align}

Finally, the results can be exploited to extract information about the LES even with nonfunctional pixels.
As elaborated in the introduction, averaging over the LES, which includes nonfunctional pixels, lead
to an underestimation of the actual performance of the LES.  In order to quantify the improvement of the presented analysis, the mean luminance of the \emph{Reference state} (left in Figure \ref{fig: reference_images_array_level_benchmark}) is determined: $\overline{L}_{\mathrm {ref}}=\left(286.4 \pm 1.1\right)\times 10^4 \, \frac{\mathrm{cd}}{\mathrm{m}^2}$.
The mean luminance analysis for all \emph{Pixel maps} yields the following:
\begin{align}
  \label{eq: mean_luminance_raw_and_denoised_pixel_mathrms}
  \begin{aligned}
    \overline{}{L}_{\mathrm {1, raw}} &= \left(256.7 \pm 0.8\right) \times 10^4 \, \frac{\mathrm{cd}}{\mathrm{m}^2}, \quad
    \overline{L}_{\mathrm {1, dn}}   = \left(275.2 \pm 0.6\right) \times 10^4 \, \frac{\mathrm{cd}}{\mathrm{m}^2} \\
    \overline{L}_{\mathrm {2, raw}} &= \left(257.0 \pm 0.8\right) \times 10^4 \, \frac{\mathrm{cd}}{\mathrm{m}^2}, \quad
    \overline{L}_{\mathrm {2, dn}}  = \left(273.9 \pm 0.5\right) \times 10^4 \, \frac{\mathrm{cd}}{\mathrm{m}^2} \\
    \overline{L}_{\mathrm {3, raw}} &= \left(258.2 \pm 0.8\right) \times 10^4 \, \frac{\mathrm{cd}}{\mathrm{m}^2}, \quad
    \overline{L}_{\mathrm {3, dn}}  = \left(275.5 \pm 0.6\right) \times 10^4 \, \frac{\mathrm{cd}}{\mathrm{m}^2}.
  \end{aligned}
\end{align}
The index ${raw}$ indicates the mean luminance, including the defect pixels,
whereas the index ${dn}$ marks the mean luminance only for functional µLEDs.
Consequently, the proposed analysis pipeline is capable of reconstructing a more
representative value for the mean luminance compared to the classical approach.


\section*{Conclusion}

The presented analysis technique is capable of reconstructing the pixel grid with high precision, even when clusters of µLEDs are not working.
Further, the machine learning-based classification of the current pixel status also shows a high overlap with the simulated yield.
In comparison, to a classical classification approach using, for instance, an arbitrary luminance threshold, no intense parameter tuning is required.
Finally, exploiting the knowledge about the µLED status enhances the analysis of the entire LES and allows to infer the actual behavior of the LES more preciously as a classical approach.
However, since the current pipeline uses an unsupervised learner (KMeans), it could behave differently on different
µLED arrays. Swapping to a supervised learner such as a \textsc{RandomForest} could reinforce the robustness of the analysis,
however, requires the presence of a labeled data set, which is time-intense to accumulate. Moreover, being able to classify thousands of µLEDs within a single measurement also offers the possibility to study the statistical distribution of each quantity for both functional and nonfunctional µLEDs.

\section*{Methods}
\subsection*{Luminance measurement}
\begin{figure}[h!]
  \centering
  \includegraphics[width=0.6\textwidth]{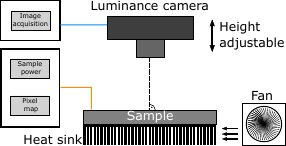}
  \caption{Schematic illustration of the used luminance setup. The height-adjustable luminance camera is aligned perpendicular to the LES. The sample is mounted on a heat sink, which is cooled actively with a fan.}
  \label{fig: schematic_experiment_setup}
\end{figure}

The used setup, shown in Figure \ref{fig: schematic_experiment_setup}, employs the luminance camera \textsc{LMK5-5} of the manufacturer TechnoTeam. The camera offers a resolution of $2448\times 2050$ pixels and is equipped with a filter wheel to consider the $V(\lambda)$ curve and color-weight functions for the measurements of luminance and color, respectively. Besides, the luminance camera uses a lens
with a magnification factor of two.
On the sample side, a white light-emitting µLED-array with a total of $60\times 60$
µLEDs, where each has size of $40 \,\mu\mathrm{m} \times 40\,\mu\mathrm{m}$, is used. Due to the luminance camera pixel's size of $3.45\,\mu\mathrm{m}$, a single µLED is represented by approximately $ 23\times 23$ camera pixels. Accordingly, the setup is theoretically capable of resolving smaller µLEDs. However, the significance of the pixel-level analysis diminishes with fewer camera pixels per µLED. For instance, local anomalies on a µLED would be averaged out with fewer camera pixels, yielding a less accurate statistical representation of the µLED array's behavior. Furthermore, the grid reconstruction reaches a resolution limit for fewer camera pixels. Tests with a narrowing lens (size was halved) indicate that the lower threshold for the algorithm lays at a single µLED representation of at least with $10 \times 10$ camera pixels, beyond this point the accuracy of the grid can diminish.

\subsection*{Image correction}
\begin{figure}[h!]
  \centering
  \includegraphics[width=0.8\textwidth]{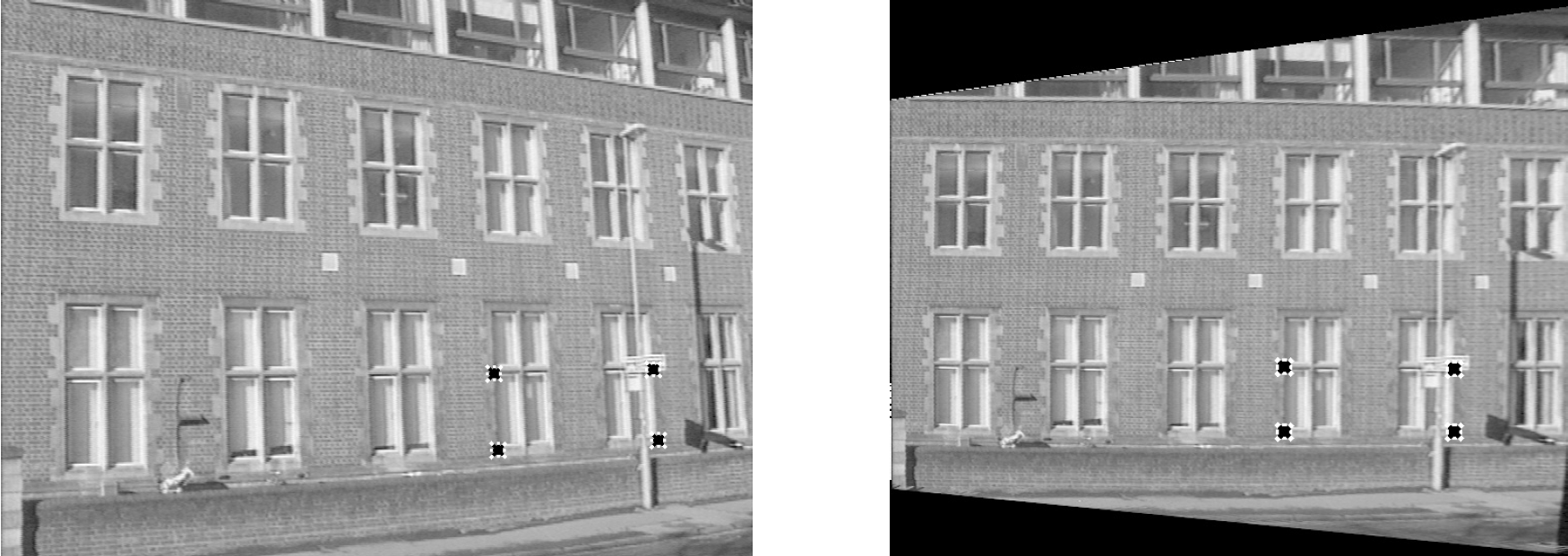}
  \caption{Illustration on how a projection transformation can remove tilting and rotation~\cite{hartley_multiple_2015}.}
  \label{fig: exemplary_outcome_of_projection_transformation}
\end{figure}

The projection transformation is a technique to project an input image
into an equivalent image, keeping its properties using a linear transformation.
A deeper insight in the mathematical description of the method is given in reference~\cite{hartley_multiple_2015}.
Figure \ref{fig: exemplary_outcome_of_projection_transformation} shows how
this transformation removes projective distortions (tilting, rotation) from an image.
Note that tilting can not only occur from a miss-aligned camera but also from the µLED array itself.
The perspective transformation is performed using \textsc{cv2.getPerspectiveTransform}
and \textsc{cv2.warpPerspective}, which are implemented in the \textsc{python} package
\textsc{cv2} (based on \textsc{OpenCV})~\cite{python_opencv_2020, bradski_opencv_2000}.
The edges of the LES are located in the luminance image by thresholding the input image
with \textsc{cv2.threshold} and then routing the output into the contour locator \textsc{cv2.findContours}.

\subsection*{Grid reconstruction}
After ensuring an adequate alignment, the pixel grid is reconstructed
by projecting the luminance image on the $\mathrm{x}$- and $\mathrm{y}$-axis, which can be formally written as:
\begin{align}
    \begin{aligned}
  L(x, y) &\in \mathbb{R}^{N\times M}, \quad \mathrm{Luminance \,\, image}\\
  \Sigma_{\mathrm{x}} L(x_i) &= \sum_{j=1}^{M} L(x_i, y_j) \,\, \forall i \in \left[1, N\right], \quad \mathrm{x-projection}\\
  \Sigma_{\mathrm{y}} L(y_j) &= \sum_{i=1}^{N} L(x_i, y_j) \,\, \forall j \in \left[1, M\right], \quad \mathrm{y-projection}.
   \end{aligned}
\end{align}

\begin{figure}[h!]
  \centering
  \includegraphics[width=0.7\textwidth]{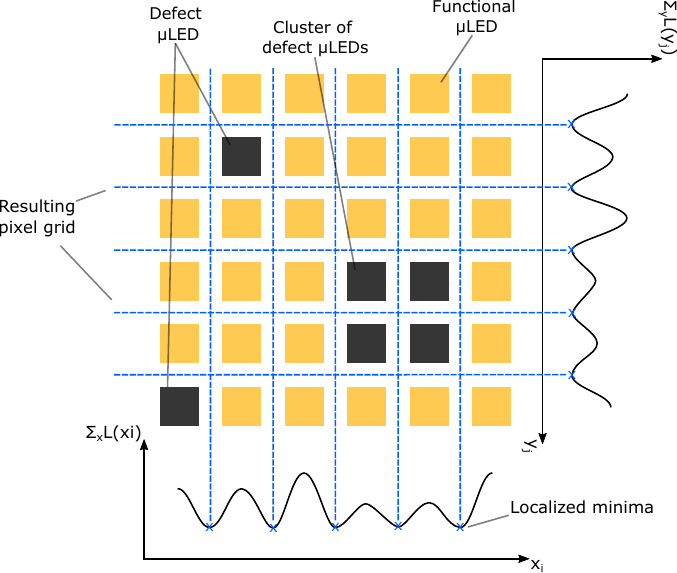}
  \caption{Illustration on how the projection of the luminance on to the $\mathrm{x}$- and $\mathrm{y}$-axis enables the reconstruction of the pixel grid.}
  \label{fig: illustration_projection_based_reconstruction}
\end{figure}

As indicated by Figure \ref{fig: illustration_projection_based_reconstruction} the values for
$\Sigma_{\mathrm{x}} L(x_i)$ and $\Sigma_{\mathrm{y}} L(y_j)$ are significantly smaller at the edge of two µLEDs, leading to minima.
Through the minima locations in the projections, it is possible to detect the pixel edges and, therefore, localize the pixel grid. Moreover, since the minima are independent of the µLEDs itself, the presented method can reconstruct the pixel grid even at positions where multiple pixels are defective.
Figure \ref{fig: reconstructed_pixel_grid_for_pixel_mathrm_1} shows the reconstructed grid for the first \emph{Pixel map}. Remarkable is that the pipeline reconstructs the pixel position even for a cluster of nonfunctional pixels correctly.

\subsection*{Pixel classification}
\begin{figure}[h!]
  \centering
  \includegraphics[width=0.5\textwidth]{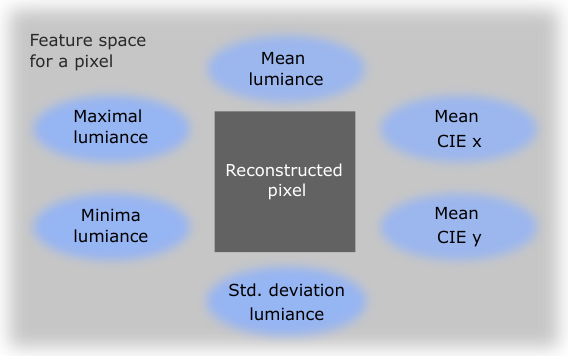}
  \caption{Visualization of the feature set for each reconstructed pixel leading to a six-dimensional parameter space.}
  \label{fig: visualizatio_feature_space}
\end{figure}

With the pixel location, a pixel-level analysis can be performed, whereby the exact type of analysis can be adapted for different use cases. In the case of the subsequent pixel classification, the mean, maximum, minimum, and standard deviation of the pixel's luminance are extracted. Additionally, the mean
color coordinates $\mathrm{CIE} \, x$, and $\mathrm{CIE} \, y$ are also calculated. In total, six parameters describe a single µLED (see Figure \ref{fig: visualizatio_feature_space}), creating a six-dimensional parameter space. This parameter space is exploited for machine learning.
Although this would be ideal for training a supervised learner such as a \textsc{Random Forest}, the lack of labeled data motivated the use of an
unsupervised learner.
From the variety of promising methods, such as \emph{Gaussian Mixture Models}, a simple \emph{k-Means-algorithm} was chosen because
it offers for this classification task a precision of over $99 \%$. In addition, its simplicity allows people with less background in machine learning to quickly implement the presented analysis pipeline.

After extracting the information shown in Figure \ref{fig: visualizatio_feature_space} for each pixel, the data is standardized with
\textsc{sklearn.preprocessing.scale}. Subsequently, a \emph{principal component analysis} (PCA) aims to increase the information density
of the feature set by reducing the number of the feature space to two. A PCA tries to find new axes in the input parameter space,
which maximize the variance of the data.
These axes correspond to the eigenvectors with the largest eigenvalues extracted from the samples
input covariance matrix. After determination of the eigenvectors, the input data is projected onto these new axes~\cite{tipping_mixtures_1999}. The reference~\cite{bishop_pattern_2009} provides a mathematical
description. On the software side, the implementation \textsc{sklearn.decomposition.PCA}
of the \textsc{python} package \textsc{scikit-learn} is applied~\cite{buitinck_api_2013}.

The outcome of the PCA is then guided into the actual classifier, which is a
\emph{k-Means-algorithm}. This algorithm tries to classify the input data into $k$
clusters by minimizing the squared distance of all data point to the cluster centers.
In the context of this paper, $k$ equals two.
A more detailed description can be found in reference~\cite{bishop_pattern_2009}. For the analysis,
the implementation \textsc{sklearn.cluster.KMeans} of \textsc{scikit-learn} with
a random seed of eight and $n_{\mathrm {init}}=100$ (number of repetitions) is utilized.
%
%
\section*{Acknowledgements}
The author thanks OSRAM Opto Semiconductors GmbH for providing the light source. In addition, the author thanks
Prof. Dr. Manfred Bayer and the chair Experimentelle Physik 2 of TU Dortmund University for supporting this work.

\section*{Disclosures}
The authors declare no conflicts of interest.

\section*{Author contributions statement}
S.B. conceived the experiments, developed and implemented the analysis, and reviewed the manuscript.

\printbibliography

\end{document}